\documentclass{elsart}

\usepackage{amssymb}
\usepackage{graphicx}
\usepackage{amsmath}
\begin{document}

\begin{frontmatter}

\title{Stabilization of a Fabry-Perot interferometer using a suspension-point
interferometer}

\author{Y. Aso,}
\ead{aso@granite.phys.s.u-tokyo.ac.jp}
\author{M. Ando,}
\author{K. Kawabe,}
\author{S. Otsuka,}
\author{K. Tsubono}
\address{Department of Physics, University of Tokyo, 7-3-1, Hongo,
 Bunkyo-ku, Tokyo, Japan.}

\begin{abstract}
A suspension-point interferometer (SPI) is an auxiliary interferometer
 for active vibration isolation, implemented at the suspension points of
 the mirrors of an interferometric gravitational wave detector.  We
 constructed a prototype Fabry-Perot interferometer equipped with an SPI
 and observed vibration isolation in both the spectrum and transfer
 function.  The noise spectrum of the main interferometer was reduced by
 40 dB below 1 Hz. Transfer function measurements showed that the SPI
 also produced good vibration suppression above 1 Hz. These results
 indicate that SPI can improve both the sensitivity and the stability
 of the interferometer.

\end{abstract}

\begin{keyword}
gravitational wave \sep active vibration isolation
 \sep suspension point interferometer
\PACS 04.80.Nn \sep 95.55.Ym
\end{keyword}
\end{frontmatter}

 \section{Introduction} 

Several large-scale laser interferometric gravitational wave detectors,
such as TAMA \cite{TAMA1}, LIGO \cite{LIGO}, GEO \cite{GEO} and VIRGO
\cite{VIRGO}, are under construction or have started scientific
observations \cite{TAMA2}\cite{TAMA3}\cite{LIGO2}\cite{GEO2}.  The main
targets of these detectors are violent astronomical events, such as
the coalescence of neutron-star binaries \cite{blair}. However, such kinds of events that
produce strong gravitational radiation are extremely rare
\cite{Kalogera}. To increase the event rate, we need to improve the
interferometers to be sensitive to events from a larger
volume of space.  At the same time, the small event rate requires us to
operate the interferometers stably over the long term in a good
condition.
 
\begin{figure}[tbp]
\begin{center}
 \includegraphics[width=9cm]{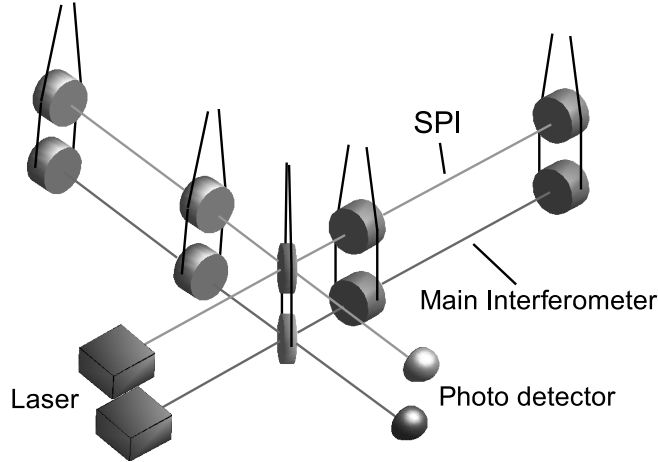}
\end{center}
\label{fpmi}
\caption{Fabry-Perot-Michelson interferometer with SPI. The lower
 interferometer (main interferometer) is used to detect gravitational waves and the
 upper one is mainly used for active vibration isolation.}
\end{figure}

A suspension-point interferometer (SPI) is an active vibration isolation
scheme that improves both the sensitivity and the stability of an
interferometer\cite{drever1}. Fig. \ref{fpmi} shows the basic
configuration of a Fabry-Perot-Michelson interferometer equipped with an
SPI. The lower interferometer is called the main interferometer and is
used to detect gravitational waves. The upper interferometer (SPI) is
mainly used for active vibration attenuation.

SPI was originally proposed by Drever more than 20 years ago as an
advanced vibration isolation scheme \cite{drever2}. He demonstrated the
stabilization of an asymmetric Michelson interferometer with corner cube
mirrors using an SPI \cite{drever3}.  However, further experimental
study on the feasibility and the practical limits of SPI, especially
using Fabry-Perot interferometers, was needed for applications of this
technique to large-scale interferometers. For this purpose, we
constructed a prototype Fabry-Perot interferometer equipped with an SPI,
strongly motivated by its possible applications to LCGT \cite{LCGT}, the
cryogenic interferometer project in Japan.

The purpose of this paper is to present results from the prototype
experiment to demonstrate the basic features of SPI.  In the next
section, we explain the working principle and the performance limits of
SPI.  We then describe the experimental setup of the prototype
interferometer, and in the following section discuss the results from the
experiment. 

\section{Suspension-Point Interferometer}
In the following section, the working principle of SPI is explained
first. We then discuss the advantages of SPI: the reduction of the residual motion,
and the low noise nature of SPI. The applications of SPI to
cryogenic interferometers and lock acquisition of interferometers
are also discussed. In section \ref{performance limits}, the theoretical
performance limits of SPI are considered.

\subsection{Working principle}\label{working principle}
\begin{figure*}[tbp]
\begin{minipage}[c]{7cm}
 \includegraphics[width=7cm]{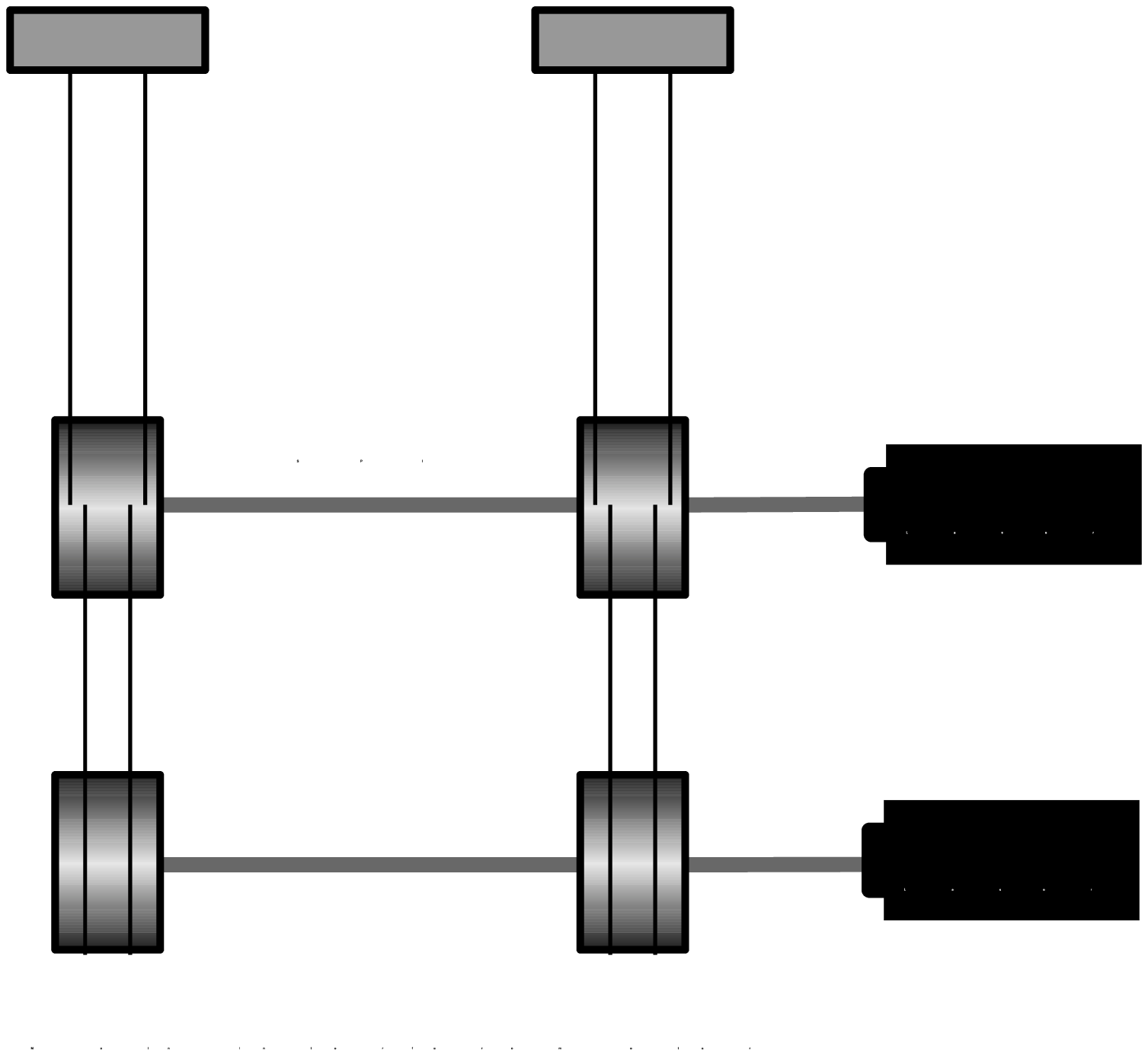}
\begin{center}
 --(a)--
\end{center}
\end{minipage}
\begin{minipage}[c]{8.4cm}
 \includegraphics[width=7cm]{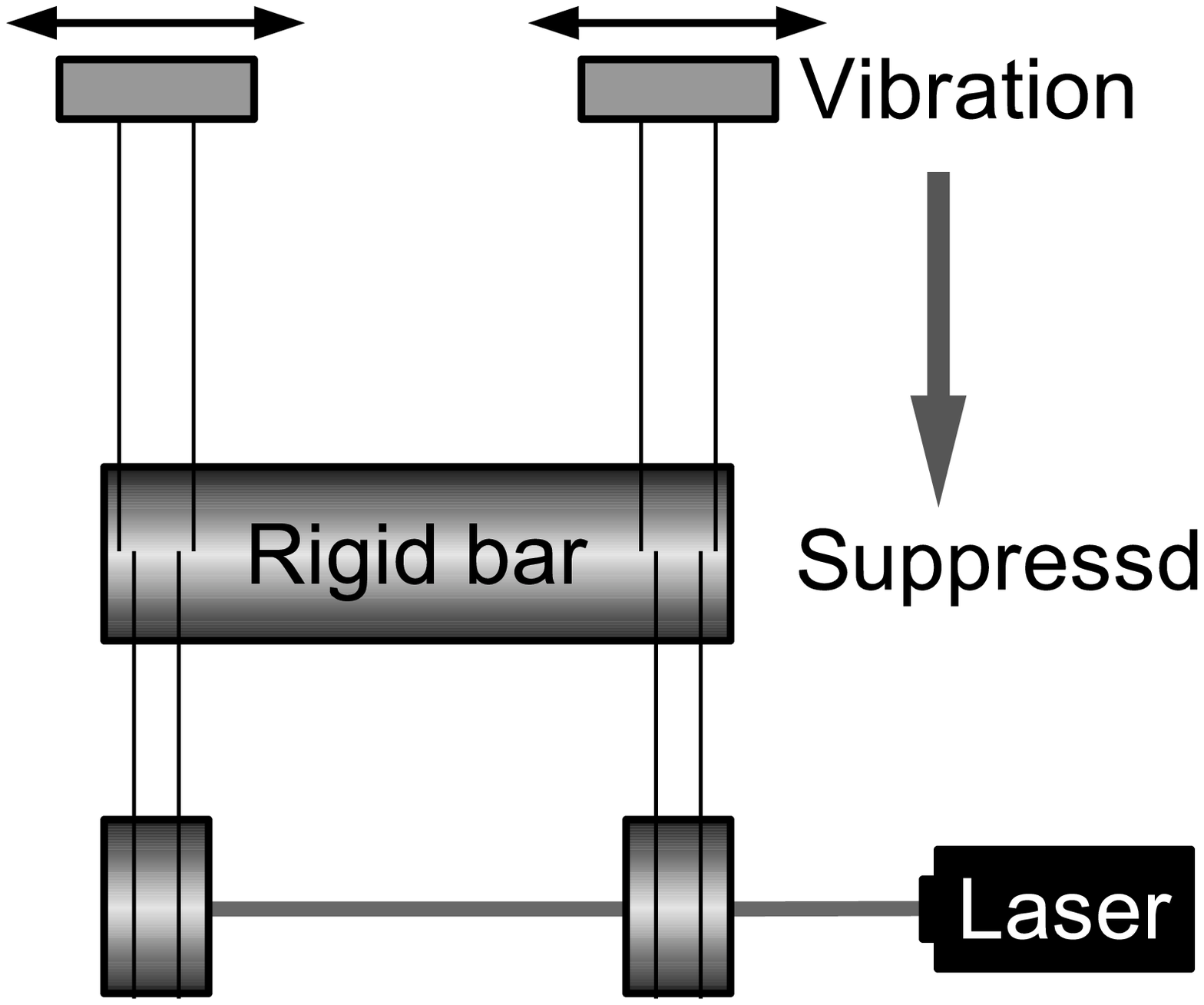}
\begin{center}
 --(b)--
\end{center}
\end{minipage}
\caption{Working principle of SPI. When the upper interferometer (SPI)
 is locked, that stage can be regarded as being a rigid bar. This rigid bar
 resists differential external disturbances.}
\label{principle}
\end{figure*}

To explain the working principle of SPI, let us first consider one arm
of a Fabry-Perot-Michelson interferometer (Fig. \ref{principle}(a)). When
we apply forces to lock the upper interferometer (SPI) to the injected
laser, the fluctuation of the distance between the two mirrors of the
SPI is suppressed by its control servo. Therefore, those mirrors form a
virtual rigid bar, as shown in Fig. \ref{principle}(b). This virtual rigid
bar resists external disturbances to change its length. Consequently, the
SPI reduces the transmission of seismic noise to the main
interferometer.  In the full configuration, as shown in Fig. \ref{fpmi},
each arm cavity is stabilized by an SPI. Consequently, the
combination of those stabilized cavities forms a stabilized
Fabry-Perot-Michelson interferometer.

In addition to the reduction of seismic noise, we can also improve the
stability of the interferometer using SPI.  When we lock the SPI, the
mirrors of the main interferometer behave as single pendulums suspended
from a rigid bar (Fig. \ref{principle}(b)). From now on we call these
single pendulums the ``main pendulums''.  This simplification of the
suspension system introduced by SPI makes it possible to suppress most
of the resonances of the complex multi-stage suspension system.  As a
result, we can reduce the residual motion of the mirrors, and thus
improve the stability of the interferometer.  Also, the smaller residual
motion allows us to use weaker coupling actuators for control of the
mirrors. This reduces the actuator noise, one of the most serious
technical noise sources in advanced interferometers.

Generally speaking, the performance of an active vibration isolation
system is limited by the noise of the sensor.  However, in the case of
SPI, the sensor is an interferometer, which potentially has the same
sensitivity as the main interferometer, itself.  Therefore, the
performance of SPI is not limited by the sensor noise. We consider
the performance-limiting factors of SPI later.

The low-noise feature of SPI is particularly useful for cryogenic
interferometers, such as LCGT. To realize a cryogenic
interferometer, we have to attach heat-link wires as close to the
mirrors as possible in order to extract heat efficiently.  However, the
heat-link wires also introduce additional seismic vibration to the
suspension stage to which the wires are attached \cite{heat link}.  To
suppress this vibration, we need a very low-noise active vibration
attenuation scheme, because any noise produced by the active system is
transmitted to the main mirrors with only a few stages of isolation.
The low-noise nature of SPI is suitable for this purpose.

SPI is also useful as a tool to help acquire the locked condition of
an interferometer, especially for the RSE (Resonant Sideband Extraction)
interferometer \cite{RSE}, which is an advanced interferometer
configuration planned to be used in next generation
interferometers. Although it is generally difficult to lock high-finesse
cavities used in RSE interferometers, reduction of the residual
motion of the mirrors introduced by SPI will make it easy to acquire the
locked condition.

\subsection{Performance limits}
\label{performance limits}
The vibration isolation ratio of SPI is mainly determined by three
factors: the control gain, common mode rejection ratio (CMRR), and coupling
from other degrees of freedom.  Using the gain $G$ of the length control
servo, the vibration attenuation factor by an SPI can be written as
$1/(1+G)$.  Usually $G$ can be set sufficiently large ($\sim 10^6$ in our
experiment) at frequencies where the seismic vibration is significant,
typically below 10Hz. Therefore, in practice the other two factors are
the dominant limitations to the performance of SPI.

First, we consider CMRR. Since Fabry-Perot interferometers are only
sensitive to the differential motion of their mirrors, the common motion
of the mirrors is not suppressed by SPI and is transmitted to the main
interferometer with the only attenuation of a single pendulum. In an
ideal case, the common motion does not produce any noise in the main
interferometer, which is also insensitive to common motion.  However, in
the real case, a fraction of the common motion of the SPI mirrors is converted
into the differential motion of the main mirrors due to asymmetries in the
main pendulums.

The CMRR of SPI is defined as the ratio of this converted differential
motion to the common motion; CMRR represents the amount of vibration
isolation produced by an SPI. Using the transfer functions from the
suspension point to the masses of the main pendulums, $H_1$ and $H_2$,
CMRR can be written as
\begin{equation}
  \text{CMRR}=2\left|\frac{H_1\left(\omega\right)-H_2\left(\omega\right)}{H_1\left(\omega\right)+H_2\left(\omega\right)}\right|.
\label{CMRR definition}
\end{equation}
In the point-mass approximation, $H_1$ and $H_2$ are explicitly written as
\begin{equation}
 H_1\left(\omega\right) =\frac{1}{1+i\frac{l_1\Gamma_1}{m_1g}\omega -\frac{l_1}{g}\omega^2},
\label{H1}
\end{equation}
\begin{equation}
H_2\left(\omega\right) =\frac{1}{1+i\frac{l_2\Gamma_2}{m_2g}\omega
 -\frac{l_2}{g}\omega^2},
\label{H2}
\end{equation}
 where $l_1$ and $l_2$ are the lengths of the main pendulums, $m_1$ and
$m_2$ are the masses of the mirrors, $\Gamma_1$ and $\Gamma_2$ are the
damping coefficients of the pendulums, and $g$ is the gravitational
constant.
Combining Eqs. (\ref{CMRR definition}),(\ref{H1}) and (\ref{H2}), we
obtain the following expression:
\begin{multline}
 \mathrm{CMRR} \simeq \\
\left|H_1\left(\omega\right)\right| \sqrt{\left(\frac{\Delta
 l}{g}\right)^2\omega^4+\left(\frac{l\Gamma}{mg}\right)^2\left(\frac{\Delta l}{l}+\frac{\Delta\Gamma}{\Gamma}-\frac{\Delta m}{m}\right)^2\omega^2},
\label{CMRR explicit form}
\end{multline}
where, $l$, $m$, and $\Gamma$ denote the mean values of the length, the
mass, and the damping coefficient of the two pendulums, respectively.  In
this approximation, only the first and second order terms of $\Delta
l \equiv l_1-l_2$, $\Delta m \equiv m_1-m_2$, and $\Delta\Gamma \equiv
\Gamma_1 -\Gamma_2$ were taken into account.

\begin{figure}[tbp]
\begin{center}
\includegraphics[width=7cm]{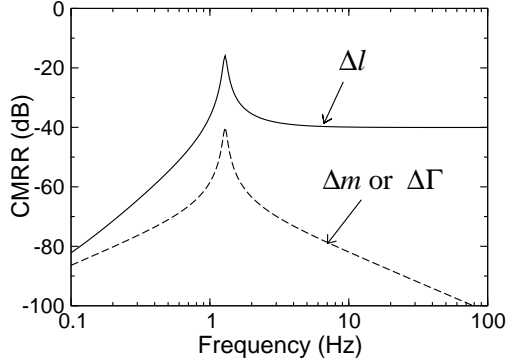}\end{center}
\caption{The CMRR of an SPI with 1\% asymmetry. The solid line shows CMRR
 with $\Delta l=1\%$. The CMRR approaches to
 $\Delta l/l$ at frequencies above 1.3Hz, which is the resonant
 frequency of the pendulums. The dashed line shows CMRR with  $\Delta
 m=1\%$, or $\Delta \Gamma =1\%$. From the symmetry of Eq. (\ref{CMRR explicit
 form}), the dependence of CMRR to $\Delta m$ and $\Delta \Gamma$ is identical.}
\label{Plot of CMRR}
\end{figure}

At frequencies higher than the resonant frequency of the main pendulums,
CMRR approaches $\Delta l/l$ (Fig. \ref{Plot of CMRR}). Therefore, the
vibration attenuation factor by SPI is limited by $\Delta l/l$ in this
frequency region.  For example, if the asymmetry of the wire length is
1\%, the limit vibration isolation ratio of the SPI is 40 dB.  Below the
resonant frequency of the main pendulums, the CMRR decreases as
the frequency goes down.

Coupling from other degrees of freedom sets another limit to the
performance of SPI.  SPI is only sensitive to the length change of the
interferometer, which is caused by the horizontal motion of the mirrors
along the direction of the laser beam. Therefore, the motions of mirrors
transverse to the beam direction, such as vertical vibration, are not
suppressed by SPI. Just like the case of CMRR, a fraction of those
motions is converted to the motion along the beam direction by
asymmetries in the suspension system and by the convergence of the local
vertical lines towards the center of the earth.  

In the case of LCGT, the amount of vertical-to-horizontal coupling
caused by the curvature of the earth is 0.2 \% \cite{arai}.  A fraction
of rotational motions is also coupled to the length fluctuation of the
interferometer, through a beam miscentering and vertical offsets of the
wire release points from the mirror's center of gravity. Considering
that the amount of those miscentering and offsets is less than 1 mm in
TAMA \cite{takamori}, the coupling coefficient from rotational motions
to the beam direction is smaller than $10^{-3}$ m/rad.  However in real
suspension systems, it is not simple to calculate the total amount of
coupling from other degrees of freedom, because it also depends on
various manufacturing imperfections. Typical measured coupling
coefficients are on the order of 1\% \cite{arai}.

\section{Experimental setup}
\begin{figure}[t]

\includegraphics[width=14.5cm]{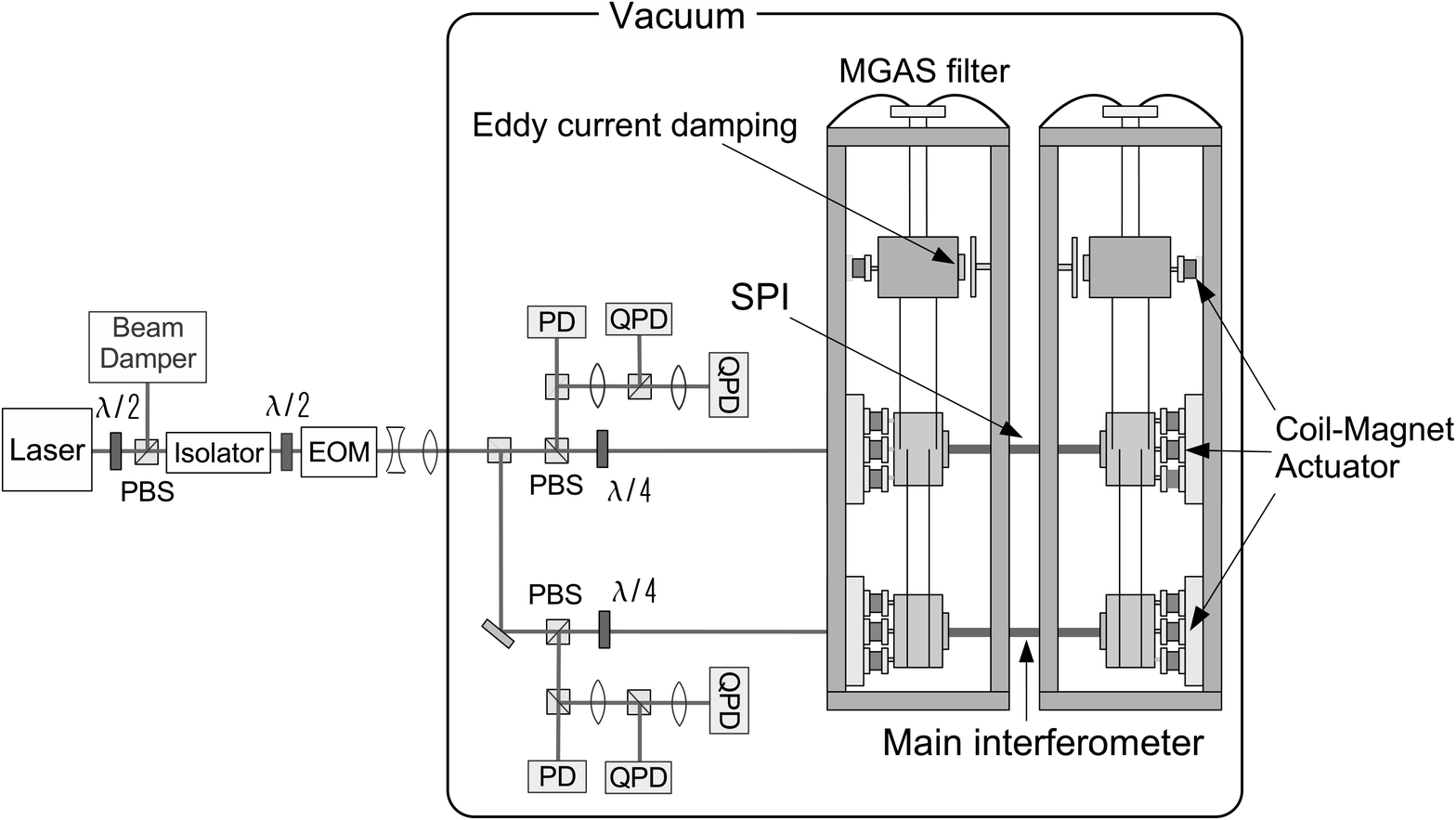}
\caption{Experimental setup. The main interferometer is a Fabry-Perot
 interferometer suspended from another Fabry-Perot interferometer
 (SPI). Light from a laser is divided by a beam splitter and injected into
 each interferometer. The reflected laser beams are detected by several
 photo-detectors to obtain the length and alignment
 signals.}\label{Setup}

\end{figure}

To test the basic features of SPI, we constructed a prototype
interferometer equipped with an SPI. As a test interferometer, we used
a Fabry-Perot interferometer instead of a full configuration
Fabry-Perot-Michelson interferometer, because it is sufficient to
test the working principle and the basic performance of SPI.

Fig. \ref{Setup} is a schematic drawing of the experimental apparatus.
The main interferometer is suspended from an SPI. Each interferometer is
a 15cm long Fabry-Perot interferometer with a finesse of 200. Our
objective is to see a reduction of the length fluctuation of the main
interferometer by locking the SPI.

The suspension system of the interferometers is a triple pendulum with
a total length of 50 cm.  The final stage and the second stage of the
suspension form the main and the suspension point Fabry-Perot
interferometers, respectively.  The masses of the first stage are called
damping masses; eddy current damping \cite{tsubono} is applied to this
stage to reduce the residual motion of the mirrors.  We used MGAS
filters \cite{MGAS}, a low-frequency vertical spring, for vertical
vibration isolation. Coil-magnet actuators are attached to each mirror
in order to control their position and orientation. We also
attached coil-magnet actuators to the damping masses to excite them and
measure the transfer function from the damping stage to the main
interferometer.

We used a 500 mW NPRO Nd:YAG laser with integrated temperature and
intensity stabilization systems. The power of the laser was attenuated
to 80 mW to avoid saturation of the photo detectors. The light emitted by
the laser is phase-modulated at 15 MHz by an EOM after a Faraday
isolator.  After passing through mode matching lenses, the light is
injected into a vacuum chamber. Inside the chamber, the beam is divided
into two beams by a beam splitter, and fed to the main and
suspension-point cavities.  The reflected light from each of the
cavities is divided by beam splitters and then detected by single-aperture
photo detectors and quadrant photo detectors. The error signals of the
length of the each cavity are obtained by the Pound-Drever-Hall (PDH)
method \cite{PDH} using signals from the single-aperture photo
detectors.  The quadrant photo detectors are placed at different Guoy
phases of the reflected beams to provide the alignment information of
the cavities with the wave-front sensing (WFS) technique
\cite{WFS1}\cite{WFS2}. Two pairs of lenses are used to adjust the Guoy
phases.

The lengths of the interferometers are locked to the laser by feeding
the PDH signals back to the coil-magnet actuators on the mirrors through
appropriate servo filters. The unity gain frequency (UGF) of the length
control servo loop is about 1 kHz in both interferometers, and the phase
margin at UGF is about 45 degrees. The open-loop gain at DC is more than
120 dB.  The interferometers are remarkably stable and continuously
locked for more than 10 hours. Actually, they have never lost the locked
state unless the servo is intentionally turned off. The alignment of the
mirrors is also controlled by feeding the WFS signals back to the
coil-magnet actuators. The bandwidth of the WFS servo is about 30
Hz. Note that the SPI works as an attenuator for the rotational as well
as longitudinal motion of the mirrors, because of this WFS servo.

\section{Results}
\begin{figure}[t]
\begin{center}
\includegraphics[width=7cm]{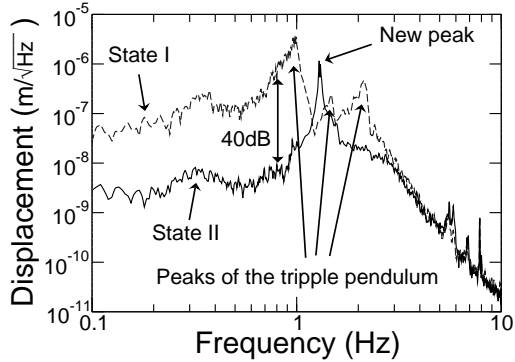} \end{center}
\caption{Equivalent displacement noise spectra of the main
 interferometer. The dashed and solid lines are the spectrum in state I
 and state II, respectively. Maximum 40 dB of noise reduction is
 observed in state II. The number of resonant peaks is reduced from
 3 in state I to 1 in state II.}
\label{spectrum}
\end{figure}

In this section we show the experimental results and compare
the two operation modes of the interferometers.  The first
operation mode is called state I, in which we only operate the main
interferometer, and the servo loop of the SPI is turned off.  The second operation mode is
called state II, in which we lock both the SPI and the main
interferometer.  In state II, the SPI works
as a vibration isolation device. We have made three types of
measurements: i) noise spectrum measurements of the main interferometer, ii)
transfer function measurements of the cavity length, iii) transfer
function measurements of the angular motion of the mirrors.

Fig. \ref{spectrum} shows the measured length fluctuation spectra of the
main interferometer between 0.1 Hz and 10 Hz. The dashed curve is the
spectrum measured in state I. The solid curve is the spectrum measured
in state II. By comparing these two curves, one can see that the
noise of the main interferometer is suppressed by the SPI below 3 Hz. The
maximum suppression ratio is 40 dB, around 1 Hz. The root mean square
(RMS) of the mirror motion integrated from 0.1 Hz to 10 Hz is
$1\times10^{-6}$m in state I, and $2\times 10^{-7}$m in the state
II. The RMS motion of state II is 5 times smaller than that of
state I.

The three peaks in state I spectrum, which correspond
to the resonances of the triple pendulum suspension, are well suppressed
by the SPI and disappear in state II.   
Instead, a sharp peak appears at 1 Hz in state II, which is the
resonance of the main pendulum. This results can be understood by the
rigid-bar picture of SPI, explained in section \ref{working principle}.  Although the
new peak in state II is very sharp, it does not make the
interferometer unstable. In fact, the peak does not appear in the
feedback signal to the actuators, because the actuation efficiency in
state II has an equally sharp peak at exactly the same frequency, and
these peaks cancel out each other.  In other words, the new peak is well
suppressed by the feedback system of the main interferometer.

\begin{figure}[t]
\begin{center}
\includegraphics[width=7cm]{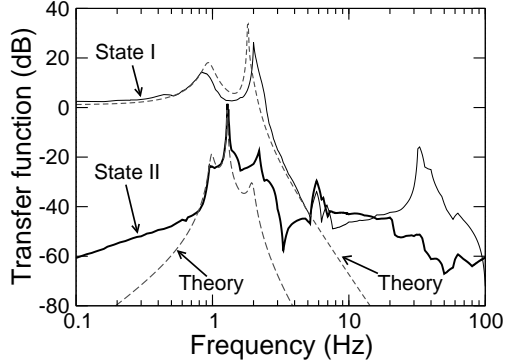}  \end{center}
\caption{Transfer functions from the displacement of a damping mass to
 the main interferometer. The damping mass was excited by means of coil magnet
 actuators and the transferred displacement was measured by the main
 interferometer. At 0.1 Hz, the vibration isolation ratio provided by SPI
 is 60 dB.  }
\label{trf}
\end{figure}

Fig. \ref{trf} shows transfer functions of the longitudinal motion from
one of the damping masses to the main interferometer. The dashed curves
are theoretical curves calculated by a simple point-mass model of the
suspension system. To calculate the theoretical curve for state II,
we adjusted the amount of the asymmetries in the model suspension
to match the measured height of the peak at 1.2 Hz.

We observed the improvement of the vibration isolation ratio by the SPI
at frequencies below 5 Hz. In the frequency region below the resonant
frequency of the main pendulums, i.e. below 1 Hz, the isolation ratio
increases as the frequency decreases. The measured behavior
qualitatively agrees with the CMRR calculation shown in Fig. \ref{Plot
of CMRR}. However, there is some discrepancy between the measurement and
the theoretical curve. The most-likely reason of this disagreement is
some kind of cross talk in the measurement system.  However, we
do not currently have any definitive explanation for this discrepancy, and further
investigation is needed.

At frequencies above 5 Hz, we could not measure the transfer function
properly due to asymmetry in the actuators on the damping mass, that
excites rotational motion of the damping mass. The ratio of the
transfered rotational motion to the horizontal motion at the main interferometer becomes
larger at higher frequencies, because the resonant frequencies of the
rotational modes of the masses are higher than that of the pendulum. The
peaks at 6 Hz and 35 Hz correspond to the yaw and pitch\footnote{``Yaw''
means rotation about the vertical axis. ``Pitch'' means rotation about
the horizontal axis which is perpendicular to the beam direction.} resonances of the main
mirror.
 Since the beam spots on the mirrors are not perfectly centered,
the rotational motion couples to the length change of the
interferometers and contaminates the transfer function measurements.
With more careful design and adjustment of the actuator balance, we should
be able to see a better improvement in the vibration isolation ratio.

\begin{figure}[t]
\begin{center}
\includegraphics[width=7cm]{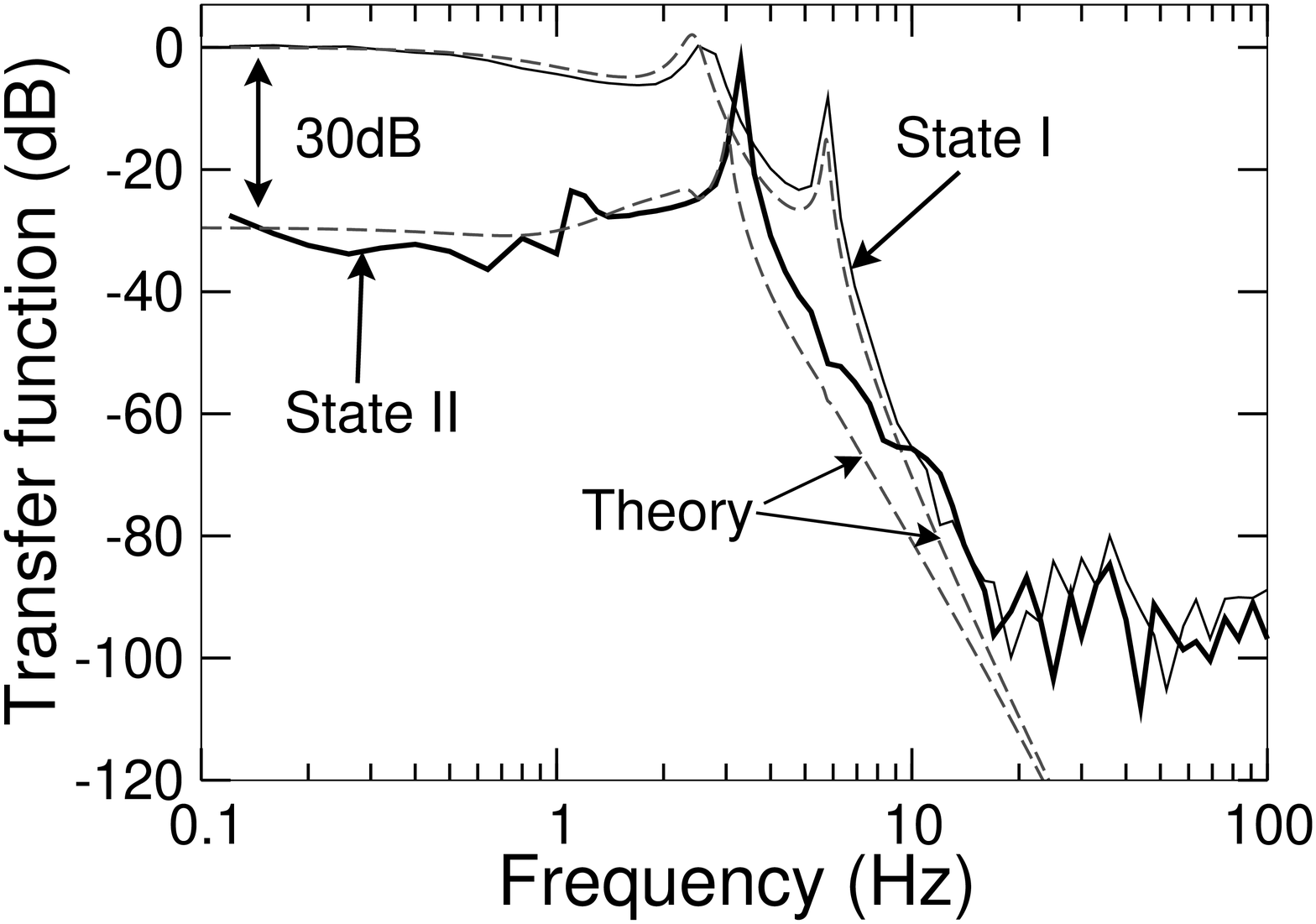} \end{center} \caption{Transfer
functions of the yaw motion from a damping mass to a mirror of the main
interferometer. The damping mass is excited by coil magnet actuators and
the yaw motion of the main mirror of the same suspension chain is
measured by the WFS signal of the main interferometer. The maximum
suppression ratio is 30dB; this is the gain of the alignment servo of
the SPI.}  \label{yaw-trf}
\end{figure}

Fig. \ref{yaw-trf} shows the transfer functions of the angular motion of
the mirrors. The yaw motion of one of the damping masses was excited by
the coil magnet actuators and the transfered yaw motion was measured by
the WFS of the main interferometer.  The theoretical curves are
calculated using a simple rigid-body model of the suspension system.

The measured transfer functions are in good agreement with the theory at
frequencies below 8 Hz. Comparing the transfer functions of state I
(the WFS servo of the SPI is off) and state II (the WFS servo of the
SPI is on), one can see that the maximum suppression ratio is 30 dB,
which corresponds to the DC gain of the WFS servo of the SPI. Above 8 Hz
the measurement is probably limited by cross talk in the electric
circuits.

\section{Discussion}
\label{discussion}
From Fig. \ref{trf}, we can see that the vibration isolation ratio of the
SPI is about 60 dB at 0.1 Hz. However, in the noise power spectrum shown
in Fig. \ref{spectrum}, the difference between state I and II is only slightly
above 20 dB at 0.1 Hz.  Besides, at frequencies above 3 Hz we can see no
difference in the noise spectrum, though the servo gain of the SPI
is still high at those frequencies. We suppose that the best
explanation for these discrepancies is vertical vibration, because
other possible reasons, such as electric noise and laser frequency
noise, are excluded by the reasoning given below.

Firstly, in the whole frequency region shown in Fig. \ref{spectrum}, the
measured electric noise was well below that of the main interferometer.
Secondly, the effective length change induced by the frequency noise is
coherent in both interferometers.  Below the resonant frequency of the
main pendulum, where the length of the main interferometer follows that
of the SPI, this effective length change of the main interferometer is
canceled by the SPI. Therefore, the frequency noise should be well
suppressed at those frequencies.  Above the resonance of the main
pendulum, the frequency noise is not canceled by the SPI, because the
coupling between the mirror motion of the two interferometers becomes
weaker. Instead, the noise of the two interferometers will be
correlated, if the dominant noise source is the frequency
noise. However, we observed no correlation between the noise of the two
interferometers at those frequencies. Hence, the dominant noise source
in this frequency range can not be the laser frequency noise. To confirm
this, we subtracted the error signals from both interferometers. If the
two signals are correlated, they should cancel out at least
partially. To determine the ratio of the subtraction, we put a
sinusoidal signal on the PZT frequency tuning of the laser to introduce
a peak in the error signals, and adjusted the ratio so that the peak
would disappear after subtraction. With this operation, no signal
cancellation was observed.  Therefore, we can exclude both electric and
frequency noises.

To obtain better performance of the SPI, we need better vertical vibration
isolation.  We have developed an active vibration suppression device for
vertical motion using a vertical suspension point
interferometer\cite{VSPI}.

Since SPI can have basically the same sensitivity as the main
interferometer, we can also use the SPI for the detection of
gravitational waves at high frequencies where seismic noise is not
significant, leaving the detection of lower frequency gravitational
waves to the main interferometer. Study on this idea is also
on going \cite{Aspen}.

\section{Summary}
A suspension point interferometer is useful for stabilization
and a sensitivity improvement of large scale interferometers.
Especially in cryogenic interferometers, SPI can be used for the
attenuation of the vibration introduced by heat link-wires. 

We have constructed and stably operated a Fabry-Perot interferometer
equipped with an SPI. Using the SPI, the length fluctuation spectrum of the main
interferometer has been reduced by 40 dB at maximum, and the RMS of the length
fluctuation has been suppressed by a factor of 5. The resonant peaks of the
triple pendulum suspension have also been suppressed by the SPI.  We believe
that we can obtain a better seismic attenuation performance with better
vertical vibration isolation.

\section{Acknowledgement}
We are grateful to Riccardo DeSalvo and Akiteru Takamori for providing
us with their MGAS filters. A part of this research is supported by a
Grant-in-Aid for JSPS Fellows and a Grant-in-Aid for Scientific Research
on Priority Areas (415) of the Ministry of Education, Culture, Sports,
Science and Technology.

\end{document}